# Image and graph convolution networks improve microbiome-based machine learning accuracy

Shtossel Oshrit [1], Isakov Haim [1], Turjeman Sondra [2],
Koren Omry [2], Louzoun Yoram [1, 3,*]

[1]Department of Mathematics, Bar-Ilan University, Ramat Gan 52900, Israel
[2] The Azrieli Faculty of Medicine, Bar-Ilan University, Safed, Israel
[3]Gonda Brain Research Center, Bar-Ilan University, Ramat Gan 52900, Israel
[*] To whom correspondences should be addressed email: louzouy@math.biu.ac.il

## 1 Abstract

The human gut microbiome is associated with a large number of disease etiologies. As such, it is a natural candidate for machine learning based biomarker development for multiple diseases and conditions. The microbiome is often analysed using 16S rRNA gene sequencing. However, several properties of microbial 16S rRNA gene sequencing hinder machine learning, including non-uniform representation, a small number of samples compared with the dimension of each sample, and sparsity of the data, with the majority of bacteria present in a small subset of samples. We suggest two novel methods to combine information from different bacteria and improve data representation for machine learning using bacterial taxonomy. iMic and gMic translate the microbiome to images and graphs respectively, and convolutional neural networks are then applied to the graph or image. We show that both algorithms improve performance of static 16S rRNA gene sequence-based machine learning compared to the best state-of-the-art methods. Furthermore, these methods ease the interpretation of the classifiers. iMic is then extended to dynamic microbiome samples, and an iMic explainable AI algorithm is proposed to detect bacteria relevant to each condition.
 Code is available at: https://github.com/oshritshtossel/iMic

## 2 Introduction

The human gut microbial composition is associated with many aspects of human health (e.g., [8, 30, 16, 33, 19, 7]). This microbial composition is often determined through sequencing of the 16S rRNA gene [58, 41]. The sequences are then clustered to produce Amplicon Sequence Variants (ASVs), which in turn are associated with bacteria [42]. This association is ambiguous, and is often not species or strain specific, but rather resolved to broader taxonomic levels (Phylum, Class, Order, Family, and Genus) [11, 45]. The 16S rRNA gene sequence-based microbial composition of a sample has been often proposed as a biomarker for disease [4, 28, 53]. Such associations can be translated to machine learning (ML) based predictions, relating the microbial composition to different conditions [36, 51, 22, 6]. However, multiple factors limit the applicability of ML in microbiome studies. First, the non-uniform representation level mentioned above requires the combination of information at different taxonomic levels. Also, in typical microbiome experiments, there are tens to hundreds of samples vs thousands of different ASVs. Finally, the ASVs are sparse, while a typical experiment can contain thousands of different ASVs. Most ASVs are absent from the vast majority of samples (See Supp. Fig. S1 for example).

To overcome these limitations, data aggregation methods have been proposed, where the hierarchical structure of the taxonomic tree can be used to combine different ASVs [45, 38]. For example, a class of phylogenetic-based feature weighting algorithms was proposed to group relevant taxa into clades, and the highly ranked clade groups in conjunction with random forest [3]. An alternative is a taxonomy-based smoothness penalty to smooth the coefficients of the microbial taxa with respect to the taxonomic tree in both linear and logistic regression models [59]. However, these simple models do not resolve the sparsity of the data and make limited use of the taxonomy.

Deep neural networks (DNNs) were proposed to identify more complex relationships among microbial taxa. Typically, the relative ASVs vectors are the input of a multi-layer perceptron neural network (MLPNN) or recursive neural network (RNN) [14]. However, given the typical distribution of microbial frequencies, these methods end up using mainly the prevalent and abundant microbes, and ignore the wealth of information available in rare bacteria.

Here, we propose to use the taxonomy to translate the microbiome to graphs or images and then apply Convolutional Neural Networks (CNNs) [26] or Graph Convolutional Networks (GCNs) [24] to the classification of such samples. CNNs have been successfully applied to diversified areas such as face recognition [39], optical character recognition [5] and medical diagnosis [52]. Convolutional Neural Networks (CNNs) are a class of neural networks that specializes in processing data with a grid-like topology, such as images [29, 43, 47, 49, 17].

Two previous models combined microbiome abundances using CNNs [47, 49]. PopPhy-CNN constructs a phylogenetic tree to preserve the relationship among the microbial taxa in the profiles. The tree is then populated with the relative abundances of microbial taxa in each individual profile and represented in a two-dimensional matrix as a natural projection of the phylogenetic tree in $R^2$. The main drawbacks of this method are the sparse image ($7\times N$, where N is the number of leaves in the tree), and the focus on a single taxonomic level and on very frequent bacteria, proven to be less useful in classification tasks [23]. Taxon-NN [49] stratifies the input ASVs into clusters based on their phylum information, then performs an ensemble of 1D-CNNs over the stratified clusters containing ASVs from the same phylum. As such, the more detailed information on the species level representation is lost.

Graphs machine learning, and specifically GCNs based graph classification tasks may also be considered for microbiome-based classification. Graph classification methods include, among many others (see also [57, 12, 31]), DIFFPOOL, a differentiable graph pooling module that can generate hierarchical representations of graphs and use this hierarchy of vertex groups to classify graphs [60]. StructPool [61] considers the graph pooling as a vertex clustering problem. EigenGCN [32] proposes a pooling operator. EigenPooling is based on the graph Fourier transform, which can utilize the vertex features and local structures during the pooling process, and QGCN [35] using a quadratic formalism in the last layer. To the best of our knowledge, GCNs have never been used for microbiome classification.

Here, we propose gMic and iMic (graph Microbiome and image Microbiome) and show that the relation between the bacteria present in a sample is often as informative as the frequency of each bacterium, and that this relation can be used to significantly improve the quality of machine learning-based prediction in microbiome-based biomarker development (micmarkers) over all existing state of the art methods. gMic and iMic address all three limitations stated above (different levels of representation, sparsity, and small number of samples).

# 3 Results

## 3.1 Taxonomic tree structure contributes to machine learning performance

Our main hypothesis is that the taxonomy tree (TT) structure of a microbiome sample is by itself an informative biomarker of the sample class, even when the frequency of each bacterium is ignored. To test this, we analysed six datasets with nine different phenotypes (Table 3 and Methods). We used 16S rRNA gene sequencing to distinguish between pathological and control cases, such as Inflammatory bowel disease (IBD), Crohn's disease (CD), Cirrhosis and different food allergies (milk, nuts and peanuts), as well as between subgroups of healthy populations by variables such as ethnicity, gender and sample type (e.g., stool vs vaginal).

We pre-processed the samples via the MIP-MLP pipeline [23]. We merged the features of the species taxonomy using the Sub-PCA method. Log normalization was used for the inputs of all the models. When species classification was unknown, we used the best-known taxonomy.

Before comparing with state-of-the-art methods comparing with the best current results, we tested three different baseline models. One was a gene frequency based naive model using a two layer, fully connected neural network. The other two models were the previous state of the art using structure, PopPhy [47], followed by one or two convolutional layers. We trained all the models on our datasets, and optimized hyperparameters for the baseline models using an NNI [2] framework on ten cross validations of the internal validation set. We measured the models' performance by their Area Under the Receiver Operator Curve (AUC). The best hyperparameters of our models were optimized using the precise same setting.

To combine the gene frequency of each bacterium through the TT, we propose gMic. We first created a TT for each dataset whose leaves are the preprocessed observed samples (each at its appropriate taxonomic level) (Fig. 1 A). The internal vertices of the TT were populated with the average over their direct descendants at the level below (e.g., for the family level, we averaged over all genera belonging to the same family). The tree was represented through a symmetric normalized adjacency matrix (see Methods). The matrix was used as the convolution kernel of a GCN. Two fully-connected layers were applied to the GCN output to predict the class of the sample. We used two versions of gMic; in the simpler version, we ignored the bacterial count, and the frequencies of all existing bacteria were replaced by a value of 1 (Fig. 1 A), and only the TT structure was used. In the second version, gMic+v, we used the normalized bacteria frequency values as the input (see Methods).

We compared the AUC on the average of the test set in 10 cross validations of gMIC and gMIC+v to the state-of-the-art results on these datasets (Fig. 1 C). The AUC, when learning using only the structure, was similar to the one of the best naive model using the gene frequencies as tabular data (see Fig. 1 C). When combined with the gene frequencies, gMIC+v outperformed existing methods in several datasets (see Table 1 and Fig.1 C).

While gMic captures the relation between similar bacteria, it still does not solve the sparsity problem. We thus suggest using iMic for a different combination of the relation between the structure of TT and the bacterial frequencies into an image and applying CNNs on this image to classify the samples (Fig. 1 B).

iMic is initiated with the same tree as gMic, but then instead of a GCN, the TT with the means in the vertices is projected to a two-dimensional matrix with seven rows (the number of different taxonomy levels in the TT), and a column for each leaf. At each level, we set the value of the matrix to be the value of the leaves, if available. Otherwise, we populate all the values at a higher level (say at the family level) to have the average values of the level below (genus). For example, if we have three genera belonging to the same family, at the family level, the three columns will receive the average of the three values at the genus level (Fig. 1 B step 3). Below a leaf, values are set to 0. As a further step to include the taxonomy in the image, columns were sorted recursively, so that similar bacteria would be closer using hierarchical clustering (see Methods).

Table 1: 10 cross validation mean performances with std on external test sets

| Model | iMic CNN2 | iMic CNN1 | gMic +v | gMic | Sub PCA | PopPhy 1 | PopPhy 2 |
|---|---|---|---|---|---|---|---|
| IBD | **0.961 ± 0.000** | 0.960±0.000 | 0.956 ± 0.02 | 0.958±0.014 | 0.94 ±0.00 | 0.863±0.02 | 0.834±0.03 |
| CD | 0.931 ±0.006 | 0.928±0.000 | **0.936 ± 0.018** | 0.870 ± 0.023 | 0.807±0.00 | 0.716 ± 0.03 | 0.751±0.016 |
| Ravel | 0.946 ± 0.01 | 0.940±0.000 | **0.977 ± 0.004** | 0.959 ± 0.006 | 0.965 ± 0.01 | 0.894 ± 0.00 | 0.898 ± 0.007 |
| Cirrhosis | **0.924±0.01** | 0.896±0.000 | 0.827±0.013 | 0.847±0.018 | 0.832±0.03 | 0.633±0.01 | 0.536±0.09 |
| Milk allergy | 0.704±0.03 | 0.704±0.04 | 0.667±0.104 | 0.707±0.04 | **0.71±0.03** | 0.557±0.04 | 0.522±0.05 |
| Nut allergy | 0.640±0.01 | **0.659±0.007** | 0.541±0.081 | 0.499±0.053 | 0.513±0.05 | 0.599±0.08 | 0.511±0.05 |
| Peanut allergy | **0.588±0.03** | 0.580±0.03 | 0.535±0.03 | 0.549±0.073 | 0.575±0.0 | 0.539±0.02 | 0.541±0.05 |
| MF | 0.641±0.06 | **0.645±0.06** | 0.446±0.04 | 0.450±0.054 | 0.51±0.102 | 0.520±0.06 | 0.527±0.06 |
| CA | **0.681±0.001** | 0.656±0.01 | 0.507±0.067 | 0.544±0.144 | 0.535±0.07 | 0.592±0.04 | 0.610±0.04 |

iMic maintains all the information (i.e., it simultaneously uses all known taxonomic levels). Moreover, it resolves the sparsity by ensuring ASVs with similar taxonomy are nearby and averaged at higher taxonomic levels. As such, even if each sample has different ASVs, there is still common information at higher taxonomic levels.

In all the datasets and all the tags above, iMic had a significantly higher average AUC than the best current methods or equal to the state of the art (one-sided T-test p value < 0.0005 on CD, nut allergy, Cirrhosis, MF and CA and p value $p < 0.05$ for milk allergy and peanut allergy (see Table 1 and Fig. 1 C).

## 3.2 Classifier interpretation

Beyond its improved performance, iMic can be used to explain the AI. We used Grad-Cam (an explainable AI platform [15]) to estimate what part of the image was used by the model to classify each class [9]. (Formally, we estimated the gradient information flowing into the last layer of the CNN to assign importance and averaged the importance of the pixels for control and case groups separately.) We show here results for the CD dataset, where we identify biomarkers to distinguish patients with CD from healthy people (Fig. 2, and Supp. Fig. S2 for another phenotype). Interestingly, the CNN is most affected by the family level (fifth row in Fig. 2), and used different bacteria for the case and the control (see Fig. 2 A, B). To find the microbes that most contributed to the classification, we projected the computed Grad-Cam values back to the TT (Fig. 2 C,D). In the CD dataset, Proteobacteria are characteristic of the CD group, in line with the literature. This phylum is proinflammatory and associated with the inflammatory state of CD and overall microbial dysbiosis [34]. Also in line with previous findings is the family *Micrococcaceae* associated with colonal CD [63] and even with mesenteric adipose tissue microbiome in CD patients [21]. The control group was characterized by *Bifidobacteriaceace*, known for its anti-inflammatory properties, pathogen resistance, and overall improvement of host state [37, 40], and by *Akkermansia*, which is a popular candidate in the search for next generation probiotics due to its ability to promote metabolism and immune function [62].

## 3.3 Sensitivity analysis

To ensure that the improved performance is not the result of hyperparameter tuning, we checked the impact on the AUC of fixing all the hyperparameters but one and changing this specific hyperparameter by increasing or decreasing its value by 10-30 percents. The difference between the AUC of the optimal parameters and all the varied combinations is low with a range of 0.005 +/- 0.015 (Supp. Mat. Fig. S3).

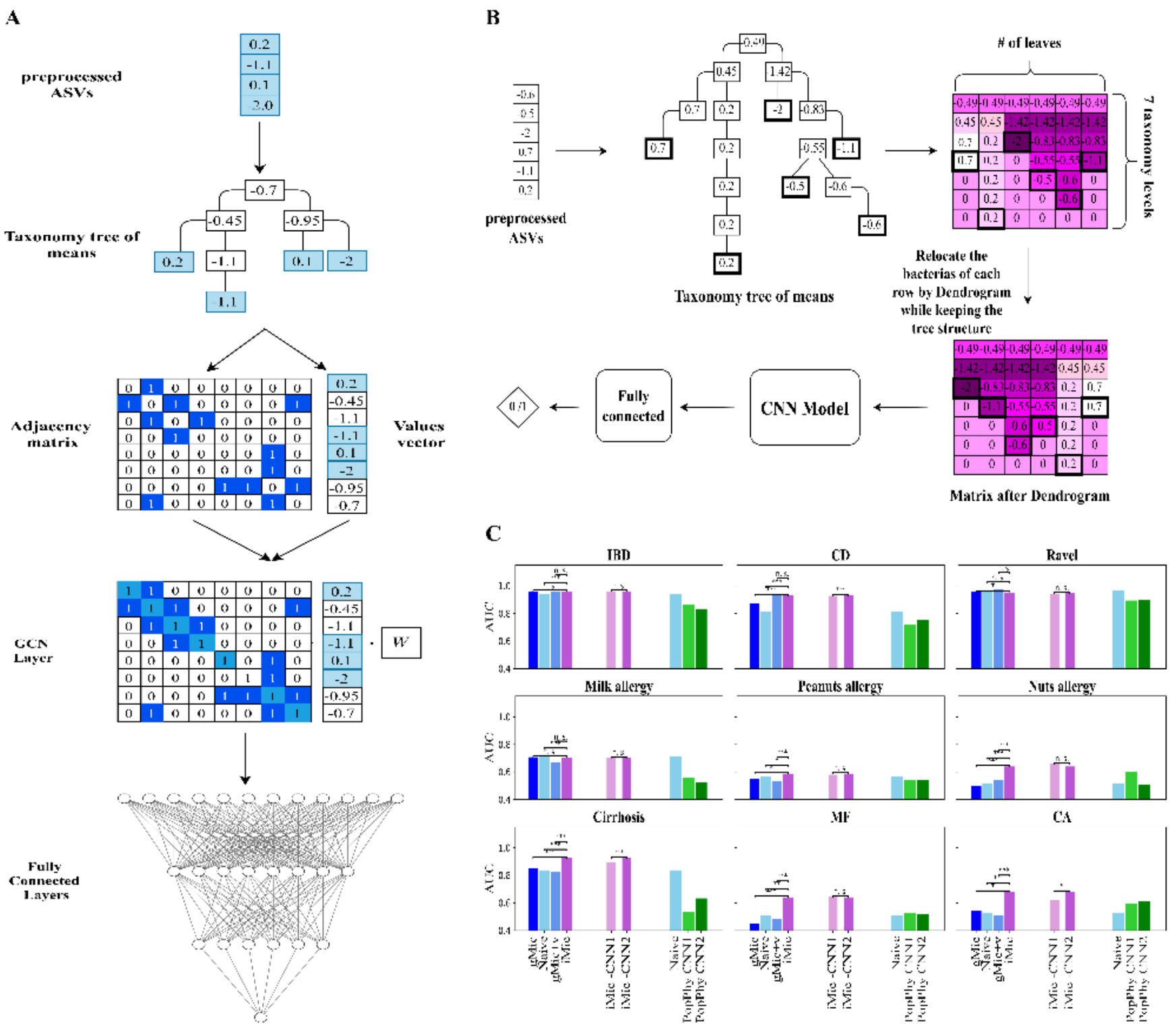

Figure 1: **iMic and gMic architectures and AUC A. gMic+v architecture:** We position all observed taxa in the leaves TT and add to each leaf its normalized frequency. Each internal node is the average of its direct descendants. These values are the input to a GCN layer with the adjacency matrix of the taxonomy tree, with the identity matrix (diagonal) added with a learned coefficient. The output of the GCN layer is fed into a fully connected layer whose output is a binary class of 0 or 1. **B. iMic architecture:** The values in the TT are as in gMic+v. The TT is then used to populate a 2-dimensional matrix. Each row in the image represents a taxonomic level. The order in each row is based on a recursive hierarchical clustering preserving the structure of the tree. The image is used

as the input of a CNN followed by 2 fully connected layers with a binary output. **C. Comparison between model performance:** The AUC is measured on the external test set on nine different phenotypes. Each subplot is a phenotype. The stars represent the significance of p value * - p=0.05, ** p=0.01, *** p=0.001. The rightmost set of plots are the baseline. The green bars are the current best baseline. The light blue bar to the right is the best baseline obtained using the MIP-MLP. The central pink bars are the iMic AUC using either a one or two dimensional CNN. The leftmost bars are for gMic (either gMic or gMic+V). We added also the iMic results to allow for a comparison.

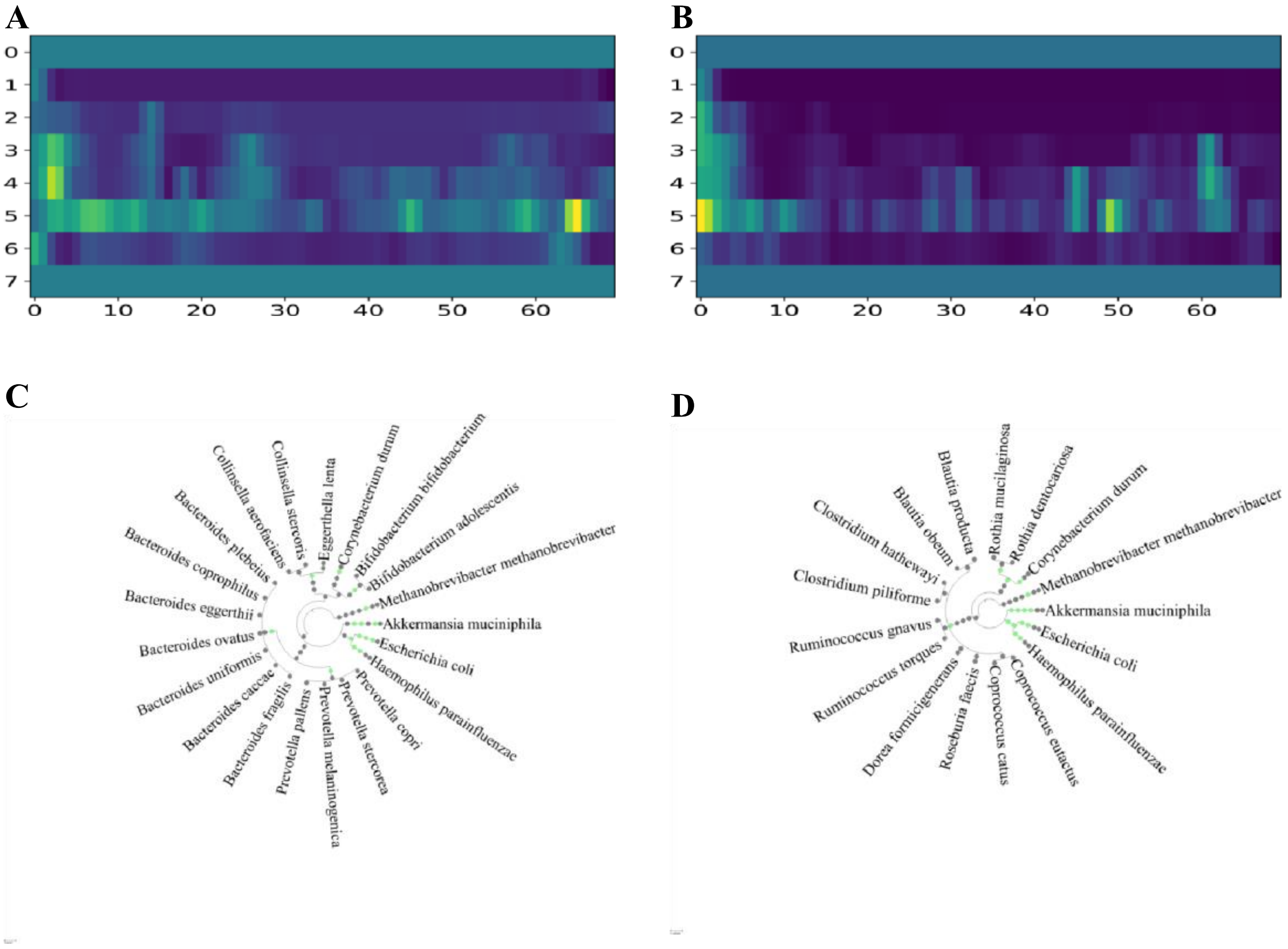

Figure 2: **Interpretation of iMic results. A, B. Grad Cam image:** Each image represents the average contribution of each input value to the gradients of the neural network back-propagation, as computed by the Grad-Cam algorithm. The results presented here are from the CD dataset. The left plot is for the healthy subjects and the right plot is for the CD subjects. The differences between the two heatmaps represent the contribution of different bacteria to the different datasets. Note that the main contribution to the classification is at the family level (row 5), suggesting that this may be the optimal level of representation. Similar results were obtained for the other datasets. **C, D. Taxonomic tree projections:** To visualize the bacteria contributing to each class, we projected the most significant microbes on the TT.

### 3.4 Temporal microbiome

iMic translates the microbiome to an image. One can use the same logic and translate a set of microbiomes to a movie to classify sequential microbiome samples. We used iMic to produce a 2-dimensional representation for the microbiome of each time step and combined those into a movie of the microbial images (see Supp. Mat. for such a movie). We used a 3D Convolutional Neural Network (3D-CNN) to classify the samples. We applied 3D-iMic to two different previously studied temporal microbiome datasets, comparing our results to the state of the art – a one dimensional representation of taxon-NN PhyloLSTM [50]. The AUC of 3D-iMic is significantly higher ($p - value < 0.0005$) than the AUC of PhyloLSTM over all datasets and tags (Fig. 3 B).

To understand what temporal features of the microbiome were used for the classification, we calculated again the heatmap of backwards gradients of each time step separately using GradCam. We focused on CNNs with a window of 3 time points and represented the heatmap of the contribution of each pixel in each time step in the R, G and B channels, producing an image that combines TT and time effects and projected this image on the TT. We used this visualization on the DiGiulio case-control study of preterm and full-term neonates'' microbiomes, and again projected the microbiome on the TT, showing the RGB representation of the contribution to the classification. Again, characteristic taxa of preterm infants (3 C, E) and full-term infants (3D ,F) were in line with previous research. Here preterm infants were characterized by TM7, common in the vaginal microbiota of women who deliver preterm ([16, 20]. Staphylococci have also been identified as main colonizers of the pre-term gut ([48, 25]). Full-term infants were characterized by a number of Fusobacteria taxa. Bacteria of this phylum are common at this stage of life [54].

# 4 Discussion

Applying machine learning to microbial gene frequencies, represented by 16S rRNA gene sequencing at a specific taxonomy level results in three types of information loss – ignoring the taxonomic relation between bacteria, ignoring sparse bacteria present in only a few samples, and ignoring rare bacteria in general. We have here proposed two novel methods (gMic and iMic) to resolve these issues and have shown that such methods produce much more precise predictions (as measured by test set AUC) than current state-of-the-art microbiome-based ML. Surprisingly, even completely ignoring the frequency of the different bacteria, and only using their absence or presence, can produce a better AUC than using just the relative frequency and ignoring the relation between bacteria. The characteristic bacterial taxa identified by our models are in line with relevant taxa previously noted in the literature.

An important advantage of iMic is the production of explainable models. Moreover, treating the microbiome as images opens the door to many vision-based machine learning tools, such as: transfer learning from pre-trained models on images, self-supervised learning, and data augmentation. The same holds for the adaptation of graph-based machine learning to microbiome graphs.

The development of microbiome-based biomarkers (micmarkers) is one of the most promising routes for easy and large-scale detection and prediction. However, while many microbiome-based prediction algorithms have been developed, they suffer from multiple limitations, which are mainly the result of the sparsity and the skewed distribution of bacteria in each host. iMic and gMic are important steps in the translation of microbiome samples from a list of single bacteria to a more holistic view of the full microbiome.

# 5 Methods

## 5.1 Preprocessing

We preprocessed the 16S rRNA gene sequences of each dataset using the MIP-MLP pipeline [23]. The preprocessing of MIP-MLP contains four stages: merging similar features based on the taxonomy, scaling the distribution, standardization to z scores, and dimension reduction. We merged the features at the species taxonomy by Sub-PCA before using all the models. We performed log normalization as well as z-scoring on the patients. No dimension reduction was used at this stage.

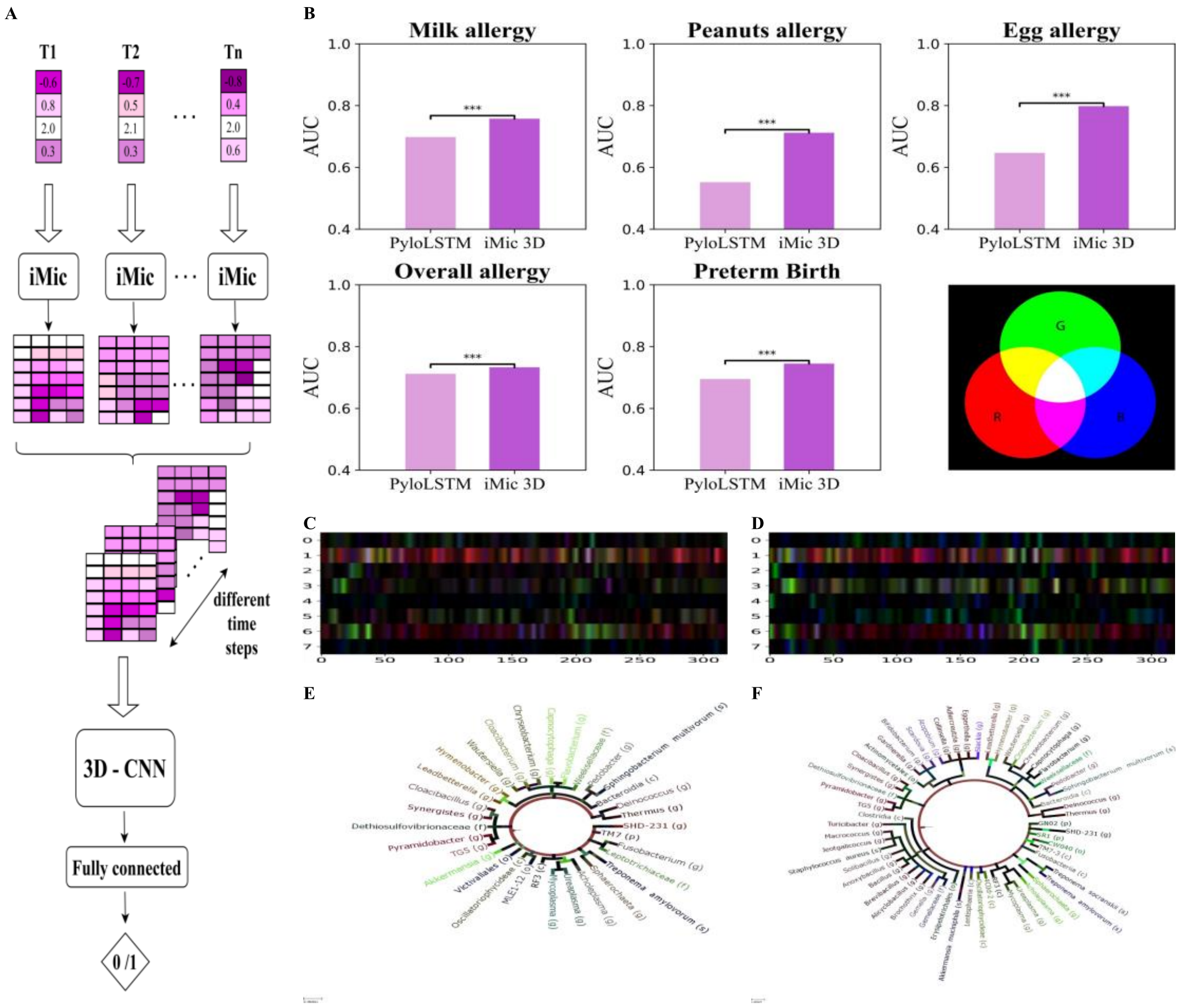

Figure 3: **3D learning: A. iMic 3D Architecture:** The ASVs frequency of each snapshot are preprocessed and combined to images as in the static iMic. The images from the different time points are combined to a 3D image, which is the input of a 3-dimensional CNN followed by two fully connected layers that return the predicted phenotype. **B. Performance of 3D learning vs PhyloLSTM:** The AUCs of the 3D-iMic are consistently higher than the AUC of the phyloLSTM on all the tags and datasets we checked. phyloLSTM is the current state of the art for these datasets (one-sided T-test, *pvalue* < 0.0005). To visualize the three-dimensional gradients (as in figure 2), we studied a CNN with a time window of 3 (i.e. 3 consecutive images are combined using the convolution). We projected the Grad Cam images to the R, G, B channels of an image. **C-D. Images after Grad Cam:** Each pixel represents the value of the backpropagated gradients after the CNN layer. The 2-dimensional image is the combination of the 3 channels above. (i.e. the gradients of the first/second/third time step are in red/green/blue). The left image is for normal birth subjects in the DiGiulio dataset, and the right image is for pre-term birth subjects. **E-F. Grad Cam projection:** Projection of the above heatmaps on the TT as in figure 2.

## 5.2 Baseline algorithms

We compared our models results to three baseline models: One gene frequency based naive model was a two layer, fully connected neural network. To evaluate the performance of using only the values. The other two models were the previous state of the art models that use structure, PopPhy [47], followed by 1 convolutional layer or 2 convolutional layers. The models' inputs were the ASVs merged at the species level by SubPCA.

### 5.2.1 Notations

To facilitate the understanding of the notations, we attach a detailed table, with all notations (Table 1).

| | |
|---|---|
| $b$ | ASVs preprocessed vector |
| $R$ | Raw representation matrix |
| $\hat{R}$ | Rearranged representation matrix |
| $l$ | A taxonomy level (Super-kingdom, Phylum, Class, Order, Family, Genus, Species) |
| $N$ | Number of means of the mean tree |
| $C_{l,i}$ | Hierarchical cluster number $i$ in level $l$ |
| $A$ | Adjacency matrix of graph |
| $\sigma$ | Activation function |
| $W$ | Weight matrix in neural network |
| $I$ | Identity matrix |
| $v$ | Gene sequence vector |

Table 2: Notations

### 5.2.2 iMic

The iMic framework consists of 3 algorithms:

- Populating the mean tree.
- Tree2Matrix.
- CNN.

Given a vector of log-normalized ASVs frequencies merged to taxonomy 7 - $b$, each entry of the vector, $b_i$ represents a microbe at a certain taxonomy level. We built an average TT, where each internal node is the average of its direct sons (see Populating mean tree algorithm and Fig. 1 B). Once the TT was populated, we built the representation matrix. We created a matrix $R \in \mathrm{R}^{7 \times N}$, where N was the number of leaves in the TT and 7 represents the 7 taxonomic levels, such that each row represents a taxonomic level.

We added the values layer by layer, starting with the values of the leaves. If there were taxonomic levels below the lead in the image, they were populated by zero. Above the leaves, we computed for each taxonomy level the average of the values in the layer below (see Fig. 1 B step 3). If the layer below had *k* different value, we set the average to all *k* positions in the current layer. For example, if there were three specie below one genera with values of 1,2 and 3. We set a value of 2 to the three positions at the genus level above these value at the specie level.

We reordered the microbes at each taxonomy to ensure that similar microbes are close to each other in the produced image. Specifically, we built a dendrogram based on the Euclidian distances as a metric using complete linkage on the rows, relocating the microbes according to the new order while keeping the phylogenetic structure. The order of the microbes was created recursively. We started by reordering the microbes on the phylum level, relocating the phylum values with all their subtree values in the matrix. Then we built a dendrogram of the descendants of each phylum separately, reordering them and their subtree in the matrix. We repeated the reordering recursively until all the microbes in the species taxonomy of each phylum were ordered. (see Reordering algorithm and see Fig. 1 B step 4).

**Algorithm 1:** Populating mean tree algorithm

**Input:** Taxonomy tree, $G = (V,E)$,a preprocessed ASVs vector, *b*
**Output:** A populated tree of means, *G*
**for** *l From the maximum tree depth to 0* **do**
    **for** *each node, v, in layer, l* **do**
        **if** *The layer of v is in b* **then**
            Assign node *v* The value from *b*
        **if** *v Has any children* **then**
            Assign its children mean to *v*
**return** *G*

**Algorithm 2:** Tree2matrix algorithm

**Input:** A populated taxonomy mean tree $G = (V,E)$
**Output:** A matrix R
Construct a zero matrix R with the number of rows equal to the layers of the tree and the number of columns equal to the number of leaves in the tree
C ← Root Node of G

```
for j from 0 to the number of layers of G do
    i ← 0
    Q ← empty Queue
    for each node v ∈ C do
        Notice: every node is a sub-tree
        if Node does not have any children then
            R(i,j) ← Abundance of node v
        else
            for kfrom1tonumberofleavesofnodev do
                R(i,j) ← Abundance of node v
                i ← i +1
            Push children of node v into queue
        i ← i +1
    C ← Q
return R
```

**Algorithm3:** Reordering algorithm, REA

```
Input: R ∈ R^{7×N}, l level of taxonomy
Output: R̂, rearranged matrix R
Target ← φ
Annotate: R in level l: b_1,b_2,...,b_N
C_{l,1},...,C_{l,k} ← Dendrogram(b_1,b_2,...,b_N)
Temp ← φ
for Cluster in C_{l,1},...,C_{l,k} do
    for Bacteria in Cluster do
        Append bacteria column to Temp if l< 7 then
            Temp ← REA (Temp,l +1)
        Append Temp to Target
return Target
```

### 5.2.3 2 dimensional CNN

The microbiome matrix was used as the input to a standard CNN [27]. We tested both one and two convolution layers. Our loss function was binary cross entropy. We used L1 regularization. We also used dropout after each layer, the strength of the dropout was controlled by a hyperparameter. For each dataset, we chose the best activation function among RelU, elU and tanh. We also used strides and padding. All the hyperparameter ranges as well as the chosen hyperparameters can be found in the Supplementary Material (Table 1 and Table 2). In order to limit the number of model parameters, we added max pooling between the layers if the number of parameters was higher than 5000. The output of the CNNs was the input of a two layer, fully connected neural network.

### 5.2.4 gMic and gMic+v

The taxonomy tree and the gene frequency vector were used as the input. The graph was represented by the symmetric normalized adjacency matrix, which was denoted $\tilde{A}$ as can be seen in the following equations:

$$\tilde{A} = D^{-\frac{1}{2}} A D^{-\frac{1}{2}} \quad (1)$$

$$D \text{ is a diagonal matrix such that } D_{ii} = \sum_j A_{ij} \quad (2)$$

The loss function was binary cross entropy. In this model, we used L2 regularization as well as drop out.

### 5.2.5 gMic

The TT was built and populated as in iMic. In gMic, a GCN layer was applied to the TT. The output of the GCN layer was the input to a fully connected neural network (FCN) as in:

$$\sigma((\tilde{A} + \alpha \cdot I) \cdot sign(v) \cdot W) \Rightarrow FCN \quad (3)$$

$sign(v)$ implies that all positive values were replaced by 1 in the gene frequency vector (i.e., the values are ignored). $\alpha$ is a learned parameter, regulating the importance given to the vertices' values against the first neighbours. The architecture of the FCN is common in all datasets. The hyperparameters may be different (for further information see Supp. Table S2): two hidden layers, each followed by an activation function (see Supp. Mat.).

### 5.2.6 gMic+v

gMic+v is equivalent to gMic, with the only exception that the positive values were not set to 1, but instead the values in the TT were used, as in iMic.

## 5.3 Data

We used nine different tags from six different datasets of 16S rRNA gene sequences to evaluate iMic and gMic+v. Four datasets are contained within the Knight Lab ML repository [1]: Cirrhosis, Caucasians from Afro Americans (CA), Male vs female (MF) and Ravel vagina. • The Cirrhosis dataset was taken from a study of 68 Cirrhosis patients and 62 healthy subjects [44].

- The MF dataset is a part of the human microbiome project (HMP) and contains 98 males and 82 females. [10].
- The CA dataset consists of 104 Caucasian and 96 Afro American vaginal samples.
- The Ravel dataset is based on the same cohort as the CA but checked another condition of the Nugent score [46]. The Nugent score is a Gram stain scoring system for vaginal swabs to diagnose bacterial vaginosis.
- The two other datasets we used are from Koren's lab. One dataset contains 137 samples with inflammatory bowel disease (IBD), including Crohn's disease (CD) and ulcerative colitis (UC), and 120 healthy samples as controls. We also used the same dataset for another task of predicting only CD from the whole society, where there were 94 CD and 163 without CD [55].

- Another dataset we used is the allergy dataset. It is a cohort of 274 people. We tried to predict three different outcomes. The first is having or not having a milk allergy, where there are 74 people with a milk allergy and 200 without. The second is having or not having a nut (walnut and hazelnut) allergy, where there are 53 with a nut allergy and 221 without. The third is having or not having an allergy to peanuts, where 79 have a peanut allergy and 195 do not [19].

We also used two sequential datasets to evaluate iMic- CNN3.

- The first dataset is the DIABIMMUNE three country cohort with food allergy outcomes (Milk, Egg, Peanut and Overall). This cohort contains 203 subjects with 7.1428-time steps on average. [56].

- The second dataset is a DiGiulio case-control study. This is a case-control study comprised of 40 pregnant women, 11 of whom delivered preterm serving as the outcome. Overall, in this study there are 3767 samples with 1420 microbial taxa from four body sites: vagina, distal gut, saliva, and tooth/gum. In addition to bacterial taxonomic composition, clinical and demographic attributes included in the dataset are gestational or postpartum day when the sample was collected, race, and ethnicity [13].

Table 3: **Table of datasets**

| Dataset | Tag | Case | Control | Origin | ASVs species | ASVs species - no zeros | Reference |
|---|---|---|---|---|---|---|---|
| **IBD** | CD or UC | 137 | 120 | Stool | 292 | 273 | [55] |
| | CD | 94 | 163 | | | | |
| **Cirrhosis** | Cirrhosis | 68 | 62 | Stool | 558 | 538 | [10] |
| **Allergy** | Milk | 74 | 200 | Stool | 8365 | 2507 | [19] |
| | Nuts | 53 | 221 | | | | |
| | Peanut | 79 | 195 | | | | |
| **CA** | Caucasian | 96 | 104 | Vagina | 529 | 348 | [46] |
| **MF** | Male | 98 | 82 | Stool | 907 | 173 | [10] |
| **Ravel** | High Nugent score | 97 | 245 | Vagina | 530 | 410 | [46] |

Table 4: **Sequential Datasets Details**

| Dataset | Tag | Case | Control | ASVs species | Range # of time steps | Ref |
|---|---|---|---|---|---|---|
| **Diabimmune** | Milk | 53 | 150 | 87 | 1- 33 | [56] |
| | Peanuts | 9 | 194 | | | |
| | Egg | 40 | 163 | | | |
| | All | 72 | 131 | | | |
| **DiGiulio case-control study** | Preterm | 11 | 29 | 321 | 3 - 158 | [13] |

## 5.4 Statistics - comparison between models

To compare between the performances of the different models, we performed a one way ANOVA test (from scipy.stats in python) on the test AUC from the 10 cross validations of all the models. Only for the models that the ANOVA test was significant, we also performed a one-sided T-test between iMic and the other models and between the two CNNs on the iMic representation (see Fig. 1 C).

## 5.5 Experimental setup

### 5.5.1 Splitting data to training set, validation set and test set

We divided the data with an external stratified test that preserves the relations between case and control into training, test, and validation sets. This ensures that the same patient cannot be simultaneously in the training and the test set. The external test was 20 percent of the whole data. The remaining 80 percent were divided into the internal validation (20 percent of the data) and the training set (60 percent).

### 5.5.2 Hyper parameters tuning

We computed the best hyperparameters for each model using a 10-fold cross validation [18] on the internal validation. We chose the hyperparameters according to the average AUC on the 10 validations. The platform we used for the optimization of the hyperparameters is (Neural Network Intelligence) NNI [2]. The hyperparameters tuned were the coefficient of the L1 loss, the weight decay, the activation function, the number of neurons in the fully connected layers, dropout, batch size and learning rate. For the CNN models we also had the kernel sizes as well as the strides and the padding as hyperparameters. The search spaces we used for each hyperparameter are: L1 coefficient was chosen uniformly from [0,1]. Weight decay was chosen uniformly from [0,0.5]. Learning rate was one of [0.001,0.01,0.05]. the batch size was [32, 64, 218, 256]. The dropout was chosen universally from [0,0.05,0.1,0.2,0.3,0.4,0.5]. We chose the best activation function from RelU, ElU and tanh. The number of neurons was chosen relatively to the input dimension. The first linear division factor from the input size was chosen randomly from [1,11]. The second layer division factor was chosen from [1,6]. The kernel sizes were defined by two different hyperparameters, a parameter for its length and its width. The length was in the range of [1,8] and the width was in the range of [1,20]. The strides were in the range of [1,9] and the channels were in the range of [1,16].

## Supplementary Material:

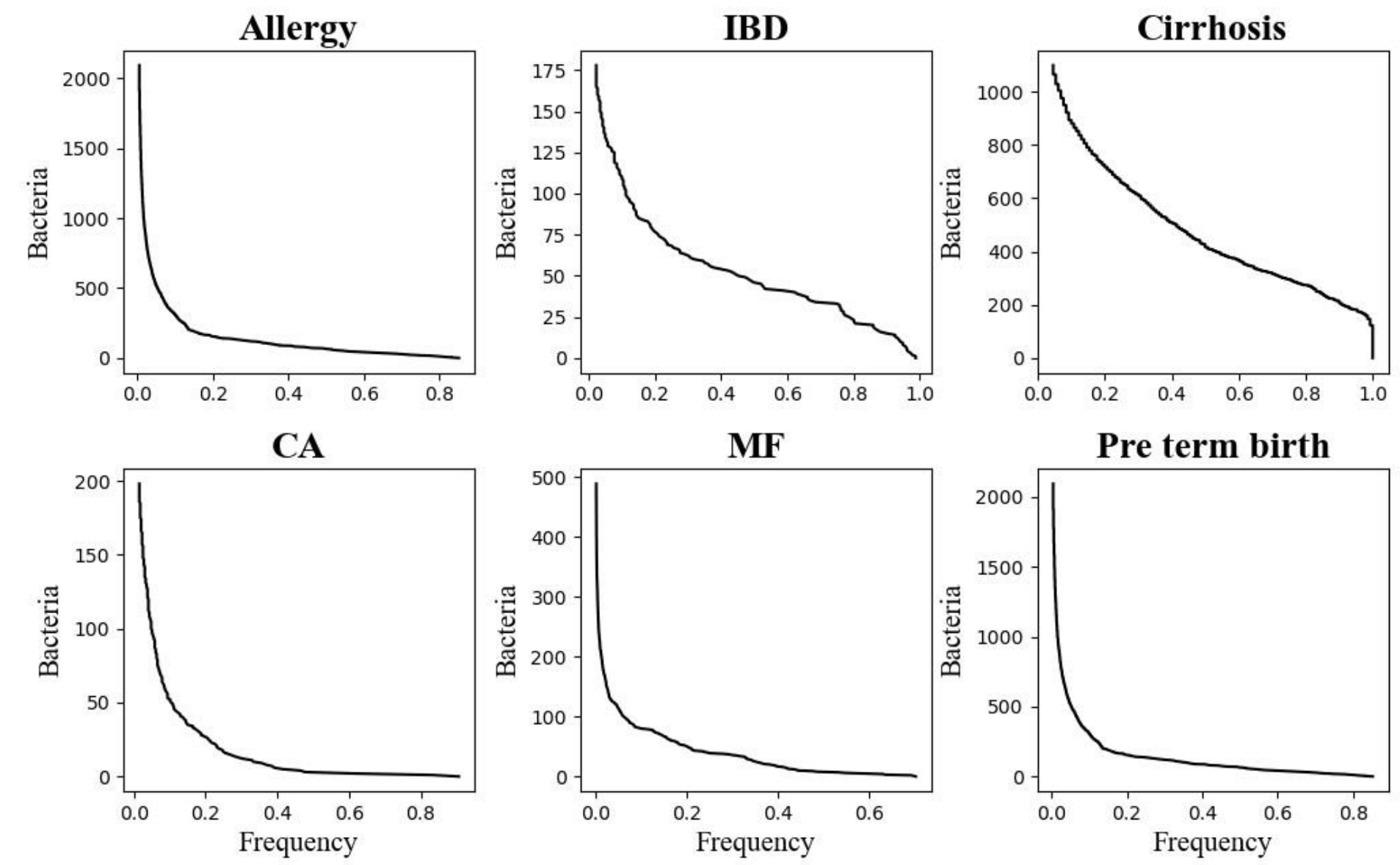

Figure S1: **Sparsity of the ASV:** Each subplot is a different dataset. The x axis is fraction of the samples in a dataset where a bacteria is found, and the y axis is the number of such bacteria. It is clear that most of the bacteria are absent from most of the datasets. The analysis was performed at the specie level.

**Links to microbiome movies:**

**Movie of DiGiulio case-control study:** https://drive.google.com/file/d/ 1qLZNOHqVolUe-

OwySmTf0mNg8egMAFJR3/view?usp=sharing

**Movie of Diabimmune case-control study:**

https://drive.google.com/file/d/1MC4KwfiIe-1ab-05GE0yNATIXsVmg0xR/view?usp=sharing

| Hyper parameter | Search space |
| --- | --- |
| L1 loss coefficient | [0,1] |
| Weight decay | [0,0.5] |
| Learning rate | [0.001,0.01,0.05] |
| Batch size | [32, 64, 218, 256] |
| Activation function | [RelU, ElU, tanh] |
| Dropout | [0,0.05,0.1,0.2,0.3,0.4,0.5] |
| Linear dimension 1 division factor | [1,11] |
| Linear dimension 2 division factor | [1,6] |
| GCN layer | [2,10] |
| Kernel size 1 of first Conv | [1,8] |
| Kernel size 2 of first Conv | [1,20] |
| Stride first Conv | [1,9] |
| **Hyper parameter** | **Search space** |
| Stride second Conv | [1,9] |
| Padding first Conv | [0,4] |
| Padding second Conv | [0,4] |
| Channels | [1,16] |
| Channel 2 | [1,16] |

Table S1: Hyper parameters search space table

| Dataset | Model | L1 loss coefficient | Weight decay | Learning rate | Batch size | Activation function | Dropout | Gcn dimension | Linear dimension 1 | Linear dimension 2 |
|---|---|---|---|---|---|---|---|---|---|---|
| **IBD** | Naive sub pca | 0.017 | 0.079 | 0.01 | 128 | tanh | 0.1 | - | 222 | 38 |
| | iMic CNN1 | 0.048 | 0.010 | 0.001 | 32 | elU | 0.4 | - | Total / 6 | Total / 5 |
| | iMic CNN2 | 0.142 | 0.039 | 0.001 | 256 | tanh | 0.2 | - | Total / 1 | Total / 8 |
| | Naive | - | 0.276 | 0.166 | 30 | Relu | 0.1 | - | 78 | 91 |
| | gMic | - | 0.021 | 0.008 | 10 | Relu | 0.1 | 5 | 124 | 121 |
| | gMic+v | - | 0.806 | 0.002 | 70 | elu | 0.15 | 9 | 67 | 23 |
| **CD** | iMic CNN1 | 0.078 | 0.094 | 0.01 | 128 | RelU | 0.2 | - | 130 | 284 |
| | iMic CNN2 | 0.154 | 0.107 | 0.001 | 64 | tanh | 0.1 | - | Total / 1 | Total / 1 |
| | Naive | 0.180 | 0.072 | 0.001 | 256 | tanh | 0.2 | - | Total / 10 | Total / 11 |
| | gMic | - | 0.006 | 0.366 | 5 | Relu | 0.06 | - | 64 | 141 |
| | gMic+v | - | 0.011 | 0.001 | 10 | elu | 0.05 | 5 | 67 | 128 |
| | gMic+v | - | 0.002 | 0.155 | 5 | elu | 0.2 | 7 | 58 | 42 |
| **Nugent** | iMic CNN1 | 0.109 | 0.04 | 0.001 | 128 | tanh | 0.2 | - | 35 | 145 |
| | iMic CNN2 | 0.225 | 0.011 | 0.001 | 64 | tanh | 0.4 | - | Total / 7 | Total / 1 |
| | Naive | 0.397 | 0.008 | 0.001 | 64 | tanh | 0 | - | 5 | 6 |
| | gMic | - | 1.7e-4 | 0.058 | 10 | Relu | 0.35 | - | 69 | 27 |
| | gMic+v | - | 0.043 | 0.257 | 5 | tanh | 0.33 | 2 | 51 | 145 |
| | gMic+v | - | 0.023 | 0.069 | 5 | tanh | 0.27 | 7 | 165 | 90 |
| **Cirrhosis** | iMic CNN1 | 0.230 | 0.002 | 0.001 | 32 | elU | 0.5 | - | 280 | 80 |
| | iMic CNN2 | 0.121 | 0.375 | 0.001 | 64 | tanh | 0.2 | - | Total / 7 | Total / 1 |
| | Naive | 0.003 | 0.092 | 0.001 | 256 | elU | 0.3 | - | Total / 9 | Total / 1 |
| | gMic | - | 0.008 | 0.004 | 70 | tanh | 0.4 | - | 86 | 50 |
| | gMic+v | - | 0.031 | 5e-4 | 5 | elu | 0.2 | 6 | 53 | 200 |
| | gMic+v | - | 0.015 | 1e-4 | 100 | tanh | 0.3 | 9 | 76 | 74 |
| **Milk allergy** | iMic CNN1 | 0.475 | 0.026 | 0.001 | 64 | tanh | 0.2 | - | 147 | 42 |
| | iMic CNN2 | 0.065 | 0.102 | 0.001 | 256 | elU | 0.3 | - | Total / 7 | Total / 4 |
| | Naive | 0.319 | 0.025 | 0.001 | 256 | elU | 0 | - | Total / 9 | Total / 1 |
| | gMic | - | 0.005 | 0.233 | 10 | Relu | 0.12 | - | 169 | 128 |
| | gMic+v | - | 0.017 | 0.122 | 5 | elu | 0.1 | 5 | 26 | 136 |
| | gMic+v | - | 0.012 | 0.051 | 5 | elu | 0.3 | 6 | 164 | 200 |
| **Nuts allergy** | iMic CNN1 | 0.277 | 0.092 | 0.001 | 128 | elU | 0.4 | - | 268 | 138 |
| | iMic CNN2 | 0.294 | 0.035 | 0.001 | 128 | tanh | 0.5 | - | Total / 8 | Total / 5 |
| | Naive | 0.398 | 0.008 | 0.001 | 64 | tanh | 0 | - | Total / 5 | Total / 6 |
| | gMic | - | 0.086 | 0.498 | 100 | elu | 0.1 | - | 113 | 81 |
| | gMic+v | - | 0.004 | 7e-4 | 5 | Relu | 0.2 | 5 | 101 | 109 |
| | gMic+v | - | 0.014 | 0.001 | 5 | tanh | 0.3 | 8 | 170 | 31 |
| **Peanuts allergy** | iMic CNN1 | 0.399 | 0.095 | 0.01 | 64 | tanh | 0.5 | - | 201 | 136 |
| | iMic CNN2 | 0.376 | 0.0007 | 0.001 | 256 | RelU | 0.05 | - | Total / 4 | Total / 4 |

| | | | | | | | | | |
|---|---|---|---|---|---|---|---|---|---|
| | Naive | 0.187 | 0.025 | 0.001 | 256 | elu | 0.5 | - | Total / 9 | Total / 3 |
| | gMic | - | 0.003 | 0.015 | 50 | tanh | 0.13 | - | 29 | 18 |
| | gMic+v | - | 0.032 | 2e-4 | 50 | tanh | 0.18 | 10 | 111 | 92 |
| | gMic+v | - | 0.005 | 0.157 | 10 | elu | 0.15 | 6 | 42 | 129 |
| **CA** | iMic CNN1 | 0.017 | 0.091 | 0.001 | 256 | RelU | 0.05 | - | 354 | 78 |
| | iMic CNN2 | 0.069 | 0.084 | 0.001 | 64 | tanh | 0.1 | - | Total / 5 | Total / 4 |
| | Naive | 0.003 | 0.092 | 0.001 | 256 | elu | 0.3 | - | Total / 9 | Total / 1 |
| | gMic | - | 0.001 | 0.014 | 5 | Relu | 0.3 | - | 168 | 122 |
| | gMic+v | - | 0.010 | 0.369 | 10 | tanh | 0.35 | 3 | 143 | 141 |
| | gMic+v | - | 0.002 | 0.003 | 5 | elu | 0.22 | 2 | 83 | 131 |
| **MF** | iMic CNN1 | 0.235 | 0.002 | 0.001 | 64 | RelU | 0.1 | - | 309 | 57 |
| | iMic CNN2 | 0.377 | 0.012 | 0.001 | 128 | RelU | 0.5 | - | Total / 9 | Total / 2 |
| | Naive | 0.029 | 0.019 | 0.001 | 128 | elu | 0.3 | - | Total / 7 | Total / 11 |
| | gMic | - | 1.5e-4 | 0.031 | 30 | elu | 0.04 | - | 61 | 180 |
| | gMic+v | - | 0.052 | 0.004 | 50 | tanh | 0.19 | 6 | 148 | 118 |
| | gMic+v | - | 0.050 | 0.001 | 5 | elu | 0.36 | 10 | 191 | 146 |

Table 2: **Table of hyper parameters used**

Figure S2: **Application of Grad-Cam to CA dataset**. The results are equivalent to figure 2, but for the CA dataset.

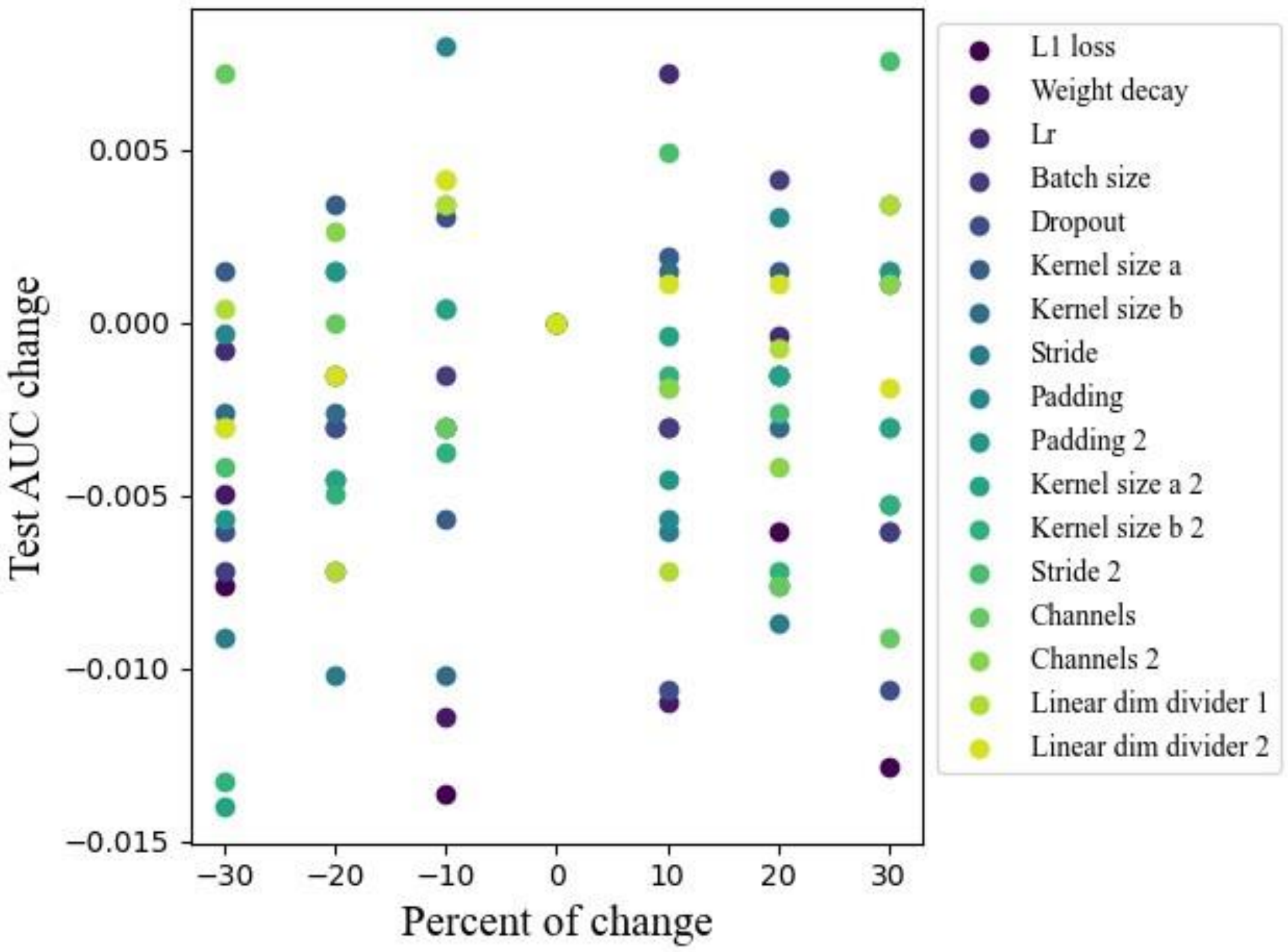

Figure S3: **Sensitivity analysis of iMic when applied to CD:** The x axis represents the change in a specific hyper parameter in percents, and the y axis represents the difference between the AUC of the hyper parameters used in Figure 1 and other combinations. The differences are low, supporting the claim that iMic is not sensitive to hyper parameters. The parameters tested are the L1 coefficient, the weight decay, the number of neurons in the fully connected layers, dropout, batch size, learning rate, and the kernel sizes as well as the strides and the padding.

Table S4: Datasets abbreviations

| Data full name | Abbrevation |
|---|---|
| Inflammatory Bowel Disease | IBD |
| Crohn's disease | CD |
| Ulcerative colitis | UC |
| Male vs female | MF |
| Caucasian vs Afro-Americans | CA |

| Dataset | Model | Kernel size 1 of first Conv | Kernel size 2 of first Conv | Kernel size 1 of second Conv | Kernel size 2 of second Conv | Stride first Conv | Stride second Conv | Padding first Conv | Padding second Conv | Channels | Channel 2 |
|---|---|---|---|---|---|---|---|---|---|---|---|
| **IBD** | iMic CNN1 | 5 | 17 | - | - | 3 | - | | | 14 | - |
| | iMic CNN2 | 2 | 5 | 2 | 4 | 2 | 3 | 1 | 3 | 9 | 14 |
| **CD** | iMic CNN1 | 5 | 10 | - | - | 3 | - | - | - | 14 | - |
| | iMic CNN2 | 3 | 6 | 2 | 5 | 1 | 2 | 3 | 2 | 4 | 16 |
| **Nugent** | iMic CNN1 | 6 | 6 | - | - | 8 | - | - | - | 15 | - |
| | iMic CNN2 | 3 | 6 | 1 | 4 | 3 | 3 | 2 | 0 | 9 | 13 |
| **Cirrhosis** | iMic CNN1 | 2 | 12 | - | - | 3 | - | - | - | 15 | - |
| | iMic CNN2 | 2 | 8 | 1 | 7 | 2 | 3 | 2 | 2 | 5 | 15 |
| **Milk allergy** | iMic CNN1 | 4 | 13 | - | - | 3 | - | - | - | 8 | - |
| | iMic CNN2 | 4 | 5 | 1 | 9 | 1 | 1 | 2 | 0 | 9 | 8 |
| **Nuts allergy** | iMic CNN1 | 3 | 11 | - | - | 5 | - | - | - | 24 | 3 |
| | iMic CNN2 | 3 | 6 | 3 | 6 | 3 | - | 2 | 0 | 9 | 13 |
| **Peanuts allergy** | iMic CNN1 | 7 | 15 | - | - | 5 | - | - | - | 12 | - |
| | iMic CNN2 | 3 | 7 | 2 | 9 | 1 | 1 | 2 | 1 | 8 | 14 |
| **CA** | iMic CNN1 | 5 | 17 | - | - | 5 | - | - | - | 9 | - |
| | iMic CNN2 | 2 | 8 | 1 | 7 | 2 | 3 | 2 | 2 | 5 | 15 |
| **MF** | iMic CNN1 | 6 | 16 | - | - | 2 | - | - | - | 13 | - |
| | iMic CNN2 | 4 | 6 | 2 | 4 | 1 | 4 | 2 | 3 | 8 | 6 |

Table S3: **Special hyper parameters of iMic**